\newcommand\lsim{\mathrel{\rlap{\lower4pt\hbox{\hskip1pt$\sim$}}
    \raise1pt\hbox{$<$}}}
\newcommand\gsim{\mathrel{\rlap{\lower4pt\hbox{\hskip1pt$\sim$}}
    \raise1pt\hbox{$>$}}}

\documentstyle[prl,aps,epsf,floats,twocolumn]{revtex}

\begin{document}
\twocolumn[\hsize\textwidth\columnwidth\hsize\csname @twocolumnfalse\endcsname
\title{The cluster abundance in cosmic string models for structure formation}

\author{
  P.P.~Avelino$^{1}$,
  J.H.P.~Wu$^{2}$,
  E.P.S.~Shellard$^{2}$
  }
\address{$^{1}$Centro de Astrof{\' \i}sica, Universidade do Porto, Rua das Estrelas s/n, 4150 Porto, Portugal}
\address{$^{2}$Department of Applied Mathematics and Theoretical Physics,
University of Cambridge,\\
Silver Street, Cambridge CB3 9EW, U.K.}

\maketitle

\begin{abstract}
We use the present observed number density of large X-ray clusters
to constrain the amplitude of matter density perturbations induced 
by cosmic strings on the scale of $8 h^{-1}$Mpc ($\sigma_8$), in both 
open cosmologies and flat models with a non-zero cosmological constant.
We find a slightly lower value of $\sigma_8$ than that obtained in the 
context of primordial Gaussian fluctuations generated during inflation. 
This lower normalization of $\sigma_8$ results from the mild non-Gaussianity 
on cluster scales, where the one point probability distribution function is
well approximated by a $\chi^2$ distribution. We use our estimate of $\sigma_8$ 
to constrain the string linear energy density $\mu$ and show that it is 
consistent with the COBE normalization.

\end{abstract}
\vskip .2in
]

\newpage

\subsection{\bf Introduction}
\label{intro}

Current theories for structure formation can be divided into two broad
categories: inflation and cosmic defects. While inflation predicts
primordial and Gaussian fluctuations
(in the simplest inflationary models), topological defects \cite{VilShe} 
induce active and non-Gaussian perturbations.

One of the most important constraints on models of structure formation is the 
the observed abundance of galaxy clusters. Although the cluster abundance 
has been widely used to constrain cosmological models with 
primordial Gaussian fluctuations (e.g. \cite{White,Eke,VL,Kitayama,Wang}) there have been few studies 
in the context of non-Gaussian perturbations such as those generated by 
topological defects. This is due to the difficulty of making 
robust predictions in topological defect scenarios,
owing to their non-linear effects which are difficult to model and require
large-scale numerical simulations  (e.g. \cite{AS1,BenBou2}). 

The present work relies on high resolution numerical simulations of cosmic 
string seeded structure formation \cite{AveShe2,AveShe4,AveShe5,AveShe3},
which are first used to estimate the power 
spectrum and one point probability distribution functions (PDF) of the 
induced density perturbations (see also 
\cite{AveCal1,against,BRA,ConHin} for different approaches). 
We then employ a simple generalization of the Press-Schechter formalism 
\cite{PS}, which is suitable for non-Gaussian perturbations with a general 
one point PDF \cite{Chiu}, in order 
to obtain the expected number density of collapsed objects with a mass 
greater that a given threshold. This generalized Press-Schechter formalism 
has been used to constrain the Gaussianity of the density fluctuations in 
the Universe \cite{Robinson2,Koyama}, and 
has been verified for a particular set of non-Gaussian structure formation 
models including a simplified flat space cosmic string model
\cite{Robinson1}.  We finally estimate the amplitude of matter density 
perturbations induced by cosmic strings on the standard scale of $8 h^{-1}$Mpc,
using the presently observed number density of large X-ray clusters. We do this for 
cosmic string models in open universes without a cosmological constant 
($\cal{S}$OCDM), and also in flat universes with a 
non-zero cosmological constant (${\cal{S}}\Lambda$CDM).
 We use our estimate of $\sigma_8$ 
to constrain the string linear energy density $\mu$ and show that it is 
consistent with the COBE normalization.

\subsection{\bf Power spectrum and PDF}
\label{powerspectrum}

In ref. \cite{AveShe3} we described the results of high-resolution 
numerical simulations of string-induced structure formation.
The power spectrum of CDM perturbations induced by long  
strings can be approximately 
described by 
\begin{equation}
  \label{Sq}
  S_{\rm CDM}(q) \equiv 4 \pi k^3 {\cal P}(k) \propto (0.7 q)^{p(q)},
\end{equation}
where
\begin{equation}
p(q) = 3.9- {{2.7} \over {1+{(2.8q)}^{-0.44}}},
\end{equation}
$q=k/\Gamma$ and the shape parameter $\Gamma$ is defined as 
\begin{equation}
  \Gamma=\Omega^0_{\rm m} h
  \exp(-2\Omega^0_{\rm B}-\Omega^0_{\rm B}/\Omega^0_{\rm m}).
\end{equation}
This fit to our numerical results has an accuracy of better 
than 10\% in the range 
$k=0.01$--$100h {\rm Mpc}^{-1}$, provided that the baryon to dark matter 
ratio is relatively small
($2\Omega^0_{\rm B}+\Omega^0_{\rm B}/\Omega^0_{\rm m} \lsim 0.3$
for 10\% accuracy in ${\cal P}(k)$).
Note, however, that there are specified uncertainties
in the underlying numerical results which are significantly 
larger \cite{AveShe3}.  This result also does not include the contribution 
from small loops which significantly enhance the overall power while 
leaving the overall shape unchanged \cite{AveShe5} (as we will discuss
later). A $\chi^2$ analysis using the observational 
power spectrum reconstructed by Peacock and Dodds \cite{PD} gives a best 
fit to long string power spectrum (\ref{Sq}) with 
$\Gamma= 0.074\times 10^{\pm 0.1}$ at the 95 per cent confidence level.
For a  baryon energy density $\Omega_B=0.05$, this implies string
models provide a consistent match to observations in the 
acceptable cosmological parameter range
$\Omega_{\rm m}^0 h = 0.14\times 10^{\pm 0.1}$.

We have used equation (\ref{Sq}) to obtain a numerical fit for the 
standard deviation of matter density perturbations at the present time as 
a function of the smoothing scale $R$ (in units of $h^{-1}{\rm Mpc}$):
\begin{equation}
  \label{sigma_R0}
  \sigma (R)=
  \sigma_{8}
  \frac{\varrho(R\Gamma)}{\varrho(8\Gamma)},
\end{equation}
where 
\begin{eqnarray}
  \varrho(r) & = & r^{-\gamma(r)},\,~~
  \gamma(r)=0.5+\frac{1.48}{1+\left(6/r\right)^{0.3}}.
  \label{sigma_n} 
\end{eqnarray}
The red-shift dependence of $\sigma$ for arbitrary $\Omega_{\rm m}$, 
$\Omega_{\Lambda}$ is described accurately by
\begin{equation}
  \label{sigma_Rz}
  \sigma (R,z)=\sigma (R)  
  \frac{g(\Omega_{\rm m},\Omega_{\Lambda})}
  {(1+z)g(\Omega^0_{\rm m},\Omega^0_{\Lambda})},
\end{equation}
where 
\begin{equation}
\label{g_O}
  g(\Omega_{\rm m},\Omega_\Lambda)  = 
  \frac{2.5\Omega_{\rm m}}
  {\left[\Omega_{\rm m}^{4/7}-\Omega_{\Lambda } +
      (1+{\Omega_{\rm m} / 2})(1+{\Omega_{\Lambda }/70})\right]}.
\end{equation}
Here, $g(\Omega_{\rm m},\Omega_\Lambda)$ gives the suppression 
of the growth of density perturbations relative to that of a 
critical-density-universe \cite{AveShe3,CarPre,AveCar}, and we
can describe the evolution of the $\Omega_{\rm m}$ with red-shift as
\begin{equation}
\label{Omegam_Z}
  \Omega_{\rm m} \equiv \Omega_{\rm m}(z)
   =  \frac{\Omega^0_{\rm m}(1+z)^3}{(1+z)^2
    (1+\Omega^0_{\rm m} z-\Omega^0_{\Lambda})+\Omega^0_{\Lambda}}.
\end{equation}
Using the simulation results for string-induced matter perturbations,
we also found
a reasonable fit to the positive side of one point PDF as follows:
\begin{equation}
  \frac{\chi^2(n)-n}{\sqrt{2n}},
\end{equation}
where $\chi^2(n)$ is the standard $\chi^2$ variant with
the number of degrees of freedom $n$, given as
\begin{equation}
\label{nR}
n=0.4 \left(R\Gamma+2\right)^{3.6}.
\end{equation}
Here $R$ is again the smoothing radius in units of 
$h^{-1}$Mpc and the fit is accurate within 10\% error for all scales. 

\subsection{\bf Modified Press Schechter formalism}
\label{Press-Schechter}

The Press-Schechter formalism \cite{PS}
relates the mass fraction of collapsed 
objects whose masses are larger than some given threshold $M$,
with the fraction 
of space in which the evolved linear density field exceeds some threshold 
$\delta_{\rm c}$. It has been extensively tested with great success 
against N-body simulations in the context of primordial 
Gaussian fluctuations, allowing for the computation of 
the number density of clusters within a given background cosmology.
Here we used an extension of the Press-Schechter formalism for 
non-Gaussian perturbations proposed by Chiu, Ostriker and Strauss \cite{Chiu} 
and verified for a particular set of non-Gaussian structure formation 
models in ref. \cite{Robinson1}. The fraction of the total mass within 
collapsed objects larger than a given mass $M$ is given by:
\begin{equation}
  {{\Omega_{\rm m}(> M(R,z),z)} \over {\Omega_{\rm m}(z)}}
  =
  f_R \times \int^\infty_{\delta_*} P_R(\nu) d \nu,
  \label{PS}
\end{equation}
where $M$ is the cluster mass defined by $M=4 \pi R^3 \rho_{\rm b}/3$, 
$f_R=1/\int_0^{\infty} P_R(\nu)d\nu$, $P_R(\nu)$ is the one point PDF
for a given smoothing radius $R$, $\nu$ is the number of standard 
deviations from mean density, $\delta_* = \delta_c / \sigma(R,z)$ and 
$\delta_{\rm c}=1.7\pm 0.2$ ($95$\% confidence interval)
assuming spherical collapse \cite{Bernardeau}.

To obtain the number density of clusters in a mass interval d$M$ about $M$
at a redshift $z$,
we differentiate the Press-Schechter formula (\ref{PS}) to obtain
(c.f. ref. \cite{VL})
\begin{equation}
  \label{ndM_PS}
  \begin{array}[l]{l}
  n(M, z){\rm d}M \approx \\
 ~~~~~ - f_R  \frac{\rho_{\rm b}}{M}
  \frac{\delta_{\rm c}}{\sigma^2(R, z)}
  \frac{{\rm d}\sigma(R,z)}{{\rm d}M}
  P_R\left[
    \frac{\delta_{\rm c}}{\sigma(R, z)}
  \right]
  {\rm d}M.
  \end{array}
\end{equation}
In the derivation of equation (\ref{ndM_PS}) we have ignored other terms 
resulting from the $R$ dependence of $P_R$ and $f_R$.
This is a mathematically motivated approximation
in the regime of mild non-Gaussianity,
and we have verified in our case that it
gives rise to at most an extra 2\% error in the final estimate of $\sigma_8$.
Substituting (\ref{sigma_Rz}) into (\ref{ndM_PS}) gives
\begin{equation}
  \label{ndM_CS}
  \begin{array}[l]{l}
  n(M, z){\rm d}M =
  f \frac{\rho_{\rm b}}{M^2}
  \frac{\delta_{\rm c}}{3\sigma(R, z)}
  P_R\left[
    \frac{\delta_{\rm c}^2}{2\sigma^2(R, z)}
  \right]\times
  \\
  ~~~~~\left\{
    \frac{0.76(R\Gamma)^{0.3}\log(R\Gamma)}{[(R\Gamma)^{0.3}+1.71]^2}
    +\gamma(R\Gamma)
  \right\}
  {\rm d}M.
  \end{array}
\end{equation}

Lacey and Cole constructed a merging history for dark matter halos based 
on the excursion set approach and obtained an analytical expression for the 
probability that a galaxy cluster with present virial mass $M$ would have
formed at some redshift $z$ \cite{LaceyCole}. The probability that a galaxy 
cluster with present virial mass $M$ would have formed at a given 
red-shift $z$ is given by:
\begin{equation}
  \label{p_z}
  p(z)=p(w(z))\frac{{\rm d}w(z)}{{\rm d}z},
\end{equation}
where
\begin{eqnarray}
  \begin{array}{ll}
    p(w(z)) & = 2 w(z)(F^{-1}-1){\rm erfc}\left({{w(z)} \over 2}\right)- \\
    & ~~~~~{\sqrt {2 \over \pi}}(F^{-1}-2)\exp\left(-{{w^2(z)} \over 2}\right),
  \end{array}
  \label{p_w_z}
  \\
  w(z) = \frac{\delta_{\rm c}(\sigma(M,0)/\sigma(M,z)-1)}
  {\sqrt{\sigma^2(FM,0)-\sigma^2(M,0)}},~~~~~
  \label{w_z}
\end{eqnarray}
and $F=0.75\pm 0.15$ ($95$\% confidence interval) is the fraction of the 
cluster mass assembled by a red-shift $z$ \cite{Navarro}. We note that this 
result was derived in the context of primordial 
Gaussian fluctuations and must 
ultimately be verified using N-body simulations.
Although it may not be valid for generic non-Gaussian models,
we still expect this to be a good approximation 
in the context of the cosmic string scenario for structure formation as the 
departures from Gaussianity on clusters scales are relatively small
\cite{AveShe4}.

\subsection{\bf The Mass-temperature relation}
\label{Mass-temperature}

In order to use the generalized Press-Schechter formalism to determine the 
abundance of X-ray clusters with a given temperature, we need to relate the 
X-ray temperature of a cluster with its virial mass. 
Here, we use the results of ref. \cite{VL} for the normalized virial 
mass-temperature relation (modified from \cite{VL}):
\begin{equation}
  \label{M_v}
  \begin{array}{l}
    M_{\rm v}=(1.23\pm0.33)\times 10^{15}
    \left[
      \frac{\Omega_{\rm mt}^{b(\Omega_{\rm mt})}}
      {\Omega_{\rm m}^0}
    \right]^{1/2}
    \left[
      \frac{1.67}{1+z_{\rm t}}\times \right. \\
    ~~~~~~~~~\left.
    {2(2-\eta)(4-\eta)^2
        \over
        64-56\eta+24\eta^2-7\eta^3+\eta^4
        }
      \frac{k_{\rm B}T}{7.5{\rm keV}}
    \right]^{3/2}
    h^{-1}{\rm M}_\odot,
  \end{array}
\end{equation}
where
$z_{\rm t}$ is the turnaround redshift,
$\Omega_{\rm mt}\equiv \Omega_{\rm m}(z_{\rm t})$,
and
\begin{eqnarray}
  \label{eta}
  \eta & \equiv & \eta(z_{\rm t}) = 
  \frac{32}{9\pi^2}
  \frac{\Omega_{\Lambda}^0 \Omega_{\rm mt}^{b(\Omega_{\rm mt})}}
  {\Omega_{\rm m}^0(1+z_{\rm t})^3},
  \\
  b(\Omega) & = & \left\{
      \begin{array}{ll}
        0.76-0.25\Omega & ({\rm OCDM}),
        \\
        0.73-0.23\Omega & ({\rm \Lambda CDM}).
      \end{array}
    \right.
\end{eqnarray}
For a given red-shift of cluster collapse $\rm z_c$,
the turnaround redshift $\rm z_t$ is easily obtained using the 
fact that $2t({\rm z_t})=t({\rm z_c})$.

Hence we can now estimate the present comoving number density of 
galaxy clusters per temperature interval d$(k_{\rm B}T)$ with a mean 
X-ray temperature $k_{\rm B}T$ which were formed at a given redshift $z$ as:
\begin{equation}
  \label{n_Tz}
  n_T(k_{\rm B}T, z){\rm d}(k_{\rm B}T){\rm d}z=
  \frac{3M}{2k_{\rm B}T}n(M,0)p(z){\rm d}(k_{\rm B}T){\rm d}z.
\end{equation}
The present abundance of X-ray clusters with a temperature
$k_{\rm B}T$ greater than $6.2 {\rm keV}$ can be estimated by 
integrating equation (\ref{n_Tz}) from $z=z_{\rm c}=0$ to infinity.
A comparison between the observed 
cluster abundance and its theoretical prediction will give an estimate 
of $\sigma_8$. 

We will use the observation for the number density of galaxy clusters 
at $z=0.05$ with X-ray temperature exceeding $6.2  {\rm keV}$,
given by Viana and Liddle \cite{VL}, based on the dataset presented by 
Henry and Arnaud \cite{Henry}, and updated by Henry \cite{Henry1}:
\begin{equation}
  N(>6.2 {\rm keV}, 0.05)
  =1.53\times 10^{-7\pm 0.16}
  h^3{\rm Mpc}^{-3}.
  \label{N_6.2_0.05}
\end{equation}
The uncertainty in (\ref{N_6.2_0.05}) is the 1-sigma interval,
and these results have taken into account the effect of
temperature measurement errors. The reasons for concentrating on 
galaxy clusters with temperature larger than $6.2 {\rm keV}$ has been extensively discussed by Viana and Liddle \cite{VL}.

\subsection{\bf Results and discussion}
\label{biaf}

\begin{figure}[t]
  \centering 
  \leavevmode\epsfxsize=8cm \epsfbox{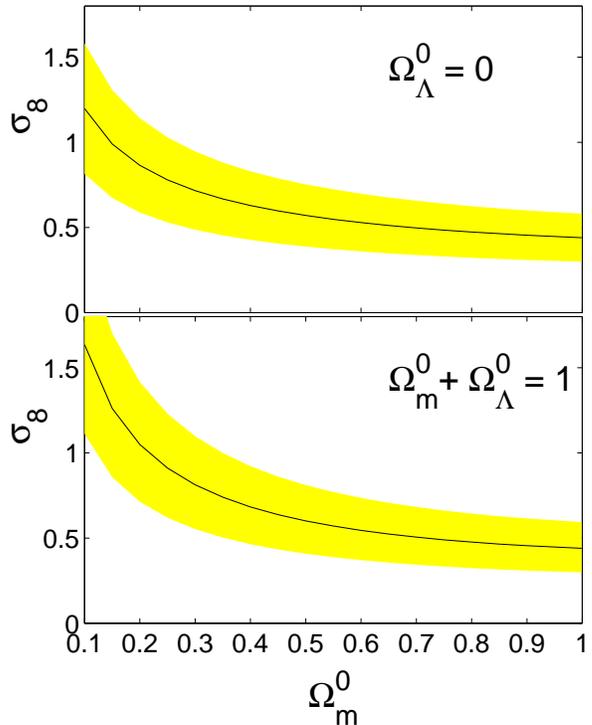}\\
  \caption[]
  {The $\sigma_{8}$ in cosmic string scenarios
    normalized to cluster abundance.
    The upper panel is the $\cal{S}$OCDM models,
    while the lower panel is the ${\cal S}\Lambda$CDM models.}
  \label{figure1}
\end{figure}

By comparing (\ref{N_6.2_0.05}) with the result integrated from (\ref{n_Tz}),
we obtain the observationally constrained $\sigma_8$ as plotted in figure
\ref{figure1}.
The overall error in the value of $\sigma_8$
was estimated by Monte Carlo simulations over $10^4$ realizations,
treating the intrinsic uncertainties in
$\Gamma$, $N(>6.2 {\rm keV}, 0.05)$ as lognormal,
and those in $\delta_{\rm c}$, $M_{\rm v}$, $F$ as Gaussian.
An accurate numerical fit to this result is
\begin{equation}
  \label{s8_fit}
  \sigma_{8}
  =
  \left\{
    \begin{array}{ll}
      0.44 \, {\Omega_{\rm m}^0}^{-0.45+0.15\Omega_{\rm m}^0}
      & \quad (\cal{S}{\rm OCDM}), \\
      0.44 \,  {\Omega_{\rm m}^0}^{-0.6+0.3\Omega_{\rm m}^0}
      & \quad ({\cal{S}}\Lambda{\rm CDM}).
    \end{array}
  \right.
\end{equation}
The $95\%$ confidence limits are
$\pm 32\%$ in the $\cal{S}$OCDM case,
and $+35\%$ and $-32\%$ in the ${\cal{S}}\Lambda$CDM case.
We note that
the overall shape of $\sigma_8$ in figure \ref{figure1} is more or less the same
as that for inflationary models \cite{VL},
while
the amplitude here is about $10$--$20\%$
lower than standard inflation.
The emergence of the same shape from these very different scenarios is due 
to the fact that 
both the string-induced and inflationary power spectra used in the calculation
of cluster abundance are constrained by the same observation \cite{PD}
and have roughly the same shape within the scales of interest
(though with quite different choices of $\Gamma$).
We have also verified that the lower normalization of $\sigma_8$ is due to
a slightly larger right-hand side tail of the PDF in the cosmic string case,
i.e.\ a high-density region then becomes consistent with 
a smaller $\sigma_8$.

The value of $\sigma_8$ in the context of the cosmic string model for
structure formation was also investigated by Bruck \cite{Bruck}. In his
work he assumed the one point PDF to be independent of
scale by taking the distribution at the non-Gaussian scale
($ \sim 1.5 (\Omega_{\rm m}^0 h^2)^{-1} {\rm Mpc}$ \cite{AveShe4})
to be valid up to scales relevant
for the cluster abundance calculation. Although this assumption can give the
right qualitative results, for small values of $\Omega_{\rm m}^0 h$, the
$R$ dependence of the one point PDF needs to be taken
into account in order to obtain more accurate results.
This improvement has been incorporated in our work,
which also took into consideration the
merging history of dark matter halos.

To see how cluster abundances constrain the string energy density per
unit length $\mu$, we can define a string ``bias'' parameter,
\begin{equation}
  \label{b_mu}
{\cal B}_{G\mu}
 (\Omega_{\rm m}^0, \Omega_{\Lambda}^0)=\frac
  {\sigma_{8}(\Omega_{\rm m}^0, \Omega_{\Lambda}^0)}
  {\sigma_{8}^{(\cal S)}(\Omega_{\rm m}^0, \Omega_{\Lambda}^0)},
\end{equation}
where $\sigma_{8}$ is the cluster result (\ref{s8_fit}) and 
$\sigma_8^{(\cal S)}$ is directly inferred 
from the long string-induced power spectrum (\ref{Sq}), given 
the assumptions that 
$\Omega^0_{\rm B}=0.05$, $\Omega_{\rm m}^0 h = 0.14$, and
$G\mu_6=G\mu\times 10^6=1$ for $\Omega_{\rm m}^0=1$ with $\Lambda=0$
(with the open and $\Lambda$-parameter dependence for $G\mu_6$ given by
$G\mu_6 \propto {\Omega_{\rm m}^0}^{-0.3}$ for ${\cal S}$OCDM
and
$G\mu_6 \propto {\Omega_{\rm m}^0}^{-0.05}$ for ${\cal S}\Lambda$CDM
\cite{AveCal1}).
\begin{figure}[t]
  \centering 
  \leavevmode\epsfxsize=8cm \epsfbox{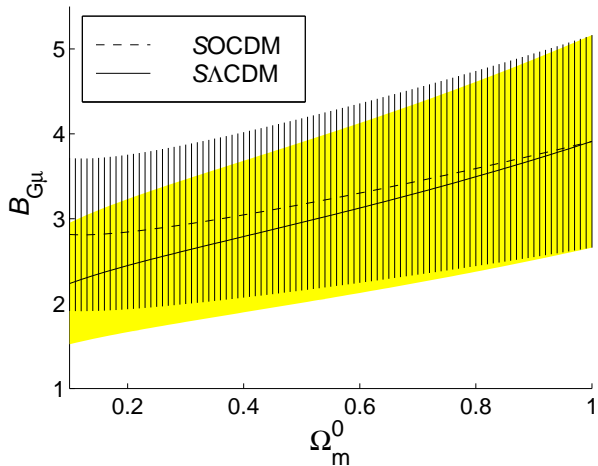}\\
  \caption[]
  {${\cal B}_{G\mu}$ as a function of $\Omega_{\rm m}^0$ and 
    $\Omega_{\Lambda}^0$.
    The hatched (${\cal S}$OCDM) and shaded (${\cal S}\Lambda$CDM)
    areas are within a 95\% confidence level.
    }
  \label{figure2}
\end{figure}

We can observe in figure \ref{figure2} for the flat $\Lambda$-model
that ${\cal B}_{G\mu}(\Omega_{\rm m}^0,\Omega_{\Lambda}^0)$
ranges from $2.2\pm 0.8$ to $3.9\pm 1.3$ (95\% confidence level)
for different choices of $\Omega_{\rm m}^0$ and $\Omega_{\Lambda}^0$.  
These values of ${\cal B}_{G\mu}$ are actually the 
cluster normalized values for $G\mu_6$ at
$\Omega_{\rm m}^0=1$ with $\Lambda=0$.
For small $\Omega_{\rm m}^0\approx 0.2\hbox{--}0.3$,
the result is slightly higher than the previously 
COBE-normalized $G\mu_6=1.7$ \cite{ACDKSS}, but it is certainly consistent 
within the uncertainties.
For large $\Omega_{\rm m}^0$, 
the result appears to be inconsistent with the COBE constraint.
However, as recently discussed in \cite{AveShe5},
the inclusion of perturbations induced by cosmic string loops 
can boost the string-induced power spectrum by
a factor of about two under reasonable assumptions.  This may serve
to remove even this apparent discrepancy in $G\mu_6$ between the COBE 
and cluster abundance
constraints.
We also note that the exact contribution 
of the background of gravitational radiation emitted by cosmic 
string loops remains a significant 
uncertainty \cite{AveCal2}.


\subsection{\bf Conclusion}
\label{conc}

We have constrained the amplitude of matter density perturbations 
induced by cosmic strings on the scale of $8 h^{-1}$Mpc in both open cosmologies 
and flat models with a non-zero cosmological constant, using the currently  
observed number density of large X-ray clusters. Because string-seeded 
matter perturbations are mildly non-Gaussian on cluster scales, we obtained a 
slightly lower normalization of $\sigma_8$ than that found for
cosmological models with primordial Gaussian fluctuations. We used the 
calculated $\sigma_8$ to constrain the string linear energy density $\mu$, 
which we found to be consistent with the latest COBE normalization when 
current uncertainties in the normalization are taken into account.


\acknowledgements

We would like to thank Pedro Viana for useful conversations.
P.\ P.\ A.\ is funded by JNICT (Portugal) under the `Program PRAXIS XXI'
(PRAXIS XXI/BPD/9901/96).
J.\ H.\ P.\ W.\ is funded by CVCP (UK) under the `ORS scheme'
(ORS/96009158) and by Cambridge Overseas Trust (UK).
This work was performed on COSMOS, the Origin2000 owned by the UK
Computational Cosmology Consortium, supported by Silicon Graphics/Cray
Research, HEFCE and PPARC.





\end{document}